\documentclass[sn-mathphys-ay]{sn-jnl}% Math and Physical Sciences Author Year Reference Style
\usepackage{graphicx}%
\usepackage{multirow}%
\usepackage{amsmath,amssymb,amsfonts}%
\usepackage{amsthm}%
\usepackage{mathrsfs}%
\usepackage[title]{appendix}%
\usepackage{xcolor}%
\usepackage{textcomp}%
\usepackage{manyfoot}%
\usepackage{booktabs}%
\usepackage{algorithm}%
\usepackage{algorithmicx}%
\usepackage{algpseudocode}%
\usepackage{listings}%

\usepackage{amsmath,esint}
\theoremstyle{thmstyleone}%
\theoremstyle{thmstyletwo}%

\theoremstyle{thmstylethree}%

\begin{document}

\title[Article Title]{A computational loudness model for electrical stimulation with cochlear implants}


\author*[1,2]{\fnm{Franklin} \sur{Alvarez}}\email{alvarez.franklin@mh-hannover.de}

\author[1,2]{\fnm{Yixuan} \sur{Zhang}}\email{zhang.yixuan@mh-hannover.de}

\author[1,2]{\fnm{Daniel} \sur{Kipping}}\email{kipping.daniel@mh-hannover.de}

\author*[1,2]{\fnm{Waldo} \sur{Nogueira}}\email{nogueiravazquez.waldo@mh-hannover.de}
%%\equalcont{These authors contributed equally to this work.}
%%\equalcont{These authors contributed equally to this work.}

\affil*[1]{\orgdiv{Department of Otorhinolaryngology}, \orgname{Hannover Medical School}, \orgaddress{\city{Hannover}, \country{Germany}}}

\affil[2]{\orgname{Cluster of Excellence ``Hearing4All''}, \orgaddress{\country{Germany}}}


\abstract{Cochlear implants (CIs) are devices that restore the sense of hearing in people with severe sensorineural hearing loss. An electrode array inserted in the cochlea bypasses the natural transducer mechanism that transforms mechanical sound waves into neural activity by artificially stimulating the auditory nerve fibers with electrical pulses. The perception of sounds is possible because the brain extracts features from this neural activity, and loudness is among the most fundamental perceptual features.

A computational model that uses a three-dimensional (3D) representation of the peripheral auditory system of CI users was developed to predict categorical loudness from the simulated peripheral neural activity. In contrast, current state-of-the-art computational loudness models predict loudness from the electrical pulses with minimal parametrization of the electrode-nerve interface. In the proposed model, the spikes produced in a population of auditory nerve fibers were grouped by cochlear places, a physiological representation of the auditory filters in psychoacoustics, to be transformed into loudness contribution. Then, a loudness index was obtained with a spatiotemporal integration over this loudness contribution. This index served to define the simulated threshold of hearing (THL) and most comfortable loudness (MCL) levels resembling the growth function in CI users.

The performance of real CI users in loudness summation experiments was also used to validate the computational model. These experiments studied the effect of stimulation rate, electrode separation and amplitude modulation. The proposed model provides a new set of perceptual features that can be used in computational frameworks for CIs and narrows the gap between simulations and the human peripheral neural activity.}




\maketitle

\section{Introduction}\label{sec_intro}

A cochlear implant (CI) is a device capable of restoring the sense of hearing in people with sensorineural hearing loss. It consists of an external device that records the sound waves that reach the outer ear, processes them, and converts them into electrical pulses that are delivered through an electrode array implanted inside the cochlea to stimulate the auditory nerve fibers (ANFs) of the peripheral auditory system \citep{Wouters2015}. Implanted subjects are often capable of understanding speech, however, there is a huge gap in performance compared to normal hearing subjects under challenging conditions \citep{Friesen2001, Nelson2004, Schorr2008, Kerber2012}. Spectrotemporal features of the neural activity are limited with electrical stimulation compared to acoustic hearing and it is in debate wether this is the cause of the performance gap \citep{Moore2003}. 

Computational models have been used to predict speech recognition performance of CI users under these challenging conditions \citep{Fredelake2012, Jurgens2018, Brochier2022, Alvarez2022}. The features used were derived from the simulated neural activity in the ANFs obtained with an electrode-nerve interface (ENI) that varied in complexity. However, there is no clear consensus on how to relate this neural activity to perceptual features in humans, therefore, the way perceptual features were extracted from this neural activity may have been very different from the way the brain is processing these inputs.

Loudness is among the most fundamental perceptual features that might be related to the neural activity along the auditory pathway \citep{Moore2003} and might help to close the gat between simulation and the reality. However, the spike count at the peripheral system has been proven not to be the only physiological correlate to loudness \citep{Carney1994, Relkin1997, Moore2003}. Instead, loudness perception is also related to the spectral loudness summation that occurs when the ANF discharges are dispersed along the basilar membrane. Recent studies suggest that the amount of neural activity at the cortex correlates better to loudness, arguably, because the auditory cortex is where the spectral loudness summation physiologically takes place \citep{Rohl2011, Rohl2012, Behler2016, Behler2021}. Additionally, \cite{Rohl2011} observed that the amount of neural activity at the brainstem correlated better to sound intensity and not to loudness, and might be the case at the peripheral level.

In psychoacoustics, the auditory filter, or critical band, is defined as the frequency bandwidth where the perception of two pure tones would interact with each other \citep{Fletcher1933, Fletcher1940}. This concept has also been used in loudness models for acoustic stimulation that account for spectral loudness summation effects. The most popular examples are the Bark scale \citep{Zwicker1957, Zwicker1980, iso2017532} and the equivalent rectangular bandwidth (ERB) \citep{Moore1983, moore1997, Glasberg2002}. However, because these auditory filters where constructed with psychoacoustic experiments, it is safe to assume that they convey neural processes that are not constrained to the peripheral system only. Existing loudness models for acoustic stimulation do not directly correlate peripheral neural activity with perceived loudness. However, they can be valuable in other aspects, which will be discussed further in this work.

McKay and colleagues have been proposing different versions of computational models to predict loudness from the biphasic pulses in electrical stimulation \citep{McKay1998, McKay2001, McKay2003, McKay2010, McKay2013a, Francart2014, Langner2020, McKay2021}. A complete review of the different variants and use cases of these models is presented by \cite{McKay2021}. For multi-electrode stimulation, McKay described a detailed loudness model (DLM) that consists of four consecutive stages: i) the temporal integration of the peripheral neural activity, ii) the transformation of neural excitation into loudness contribution, iii) the integration of loudness contributions across cochlear places to obtain the instantaneous loudness, and iv) the evaluation criterion that depended on the experiment performed. The DLM successfully reproduced the relation between the stimulation current (in dB) delivered through the electrodes and the perceived loudness by CI users. With small variations of the current at low stimulation levels, this relation can be modeled with a power function. It is argued that this power function models the changes in the peripheral neural activity with respect to the stimulation current \citep{McKay2001}. Since peripheral neural activity is concentrated near the stimulating electrode, loudness perception increases linearly with this activity. However, this relation becomes exponential with a broader stimulation range \citep{Zeng1992, Chatterjee2000, McKay2001}. This exponential relation might be caused by the spectral loudness summation when ANFs from other cochlear places are being recruited \citep{McKay2001, McKay2003}. Some of the limitations of the DLM are that the peripheral neural activity is estimated with a probabilistic model of the neuron's firing efficiency and that the definition of cochlear places is left open to interpretation \citep{McKay1998, McKay2001, McKay2013a}. The probabilistic model oversimplifies many physiological aspects compared to, for example, the before mentioned computational models to predict speech performance in CI users.

\cite{McKay2021} also presented the practical method, that was developed based on observations where the loudness contribution in sequential stimulation was not affected by the location of the electrodes \citep{McKay2003, McKay2010, McKay2021}. This was an advantage over the DLM since a definition of cochlear places was not required. With the practical method, the loudness contribution of each electrical pulse delivered through the electrode array was calculated with a previously fitted loudness growth function (LGF) for each electrode \citep{McKay2003}. This method was ideal for more complex stimuli and also changed the order of the original stages of the DLM as follows: i) the loudness contribution calculation for each electrical pulse, ii) the temporal integration of the loudness contribution, and iii) the evaluation criterion \citep{McKay2003, McKay2010, McKay2013a, Francart2014, Langner2020}. Note that the temporal integration, which was taken from models of acoustic hearing \citep{Oxenham1994, Oxenham2001}, is now applied over the loudness contribution and not over the peripheral neural activity. This might also indicate that this process occurs in higher levels of the auditory pathway.

The aim of this work was to propose a computational loudness model capable of predicting categorical loudness from the simulated peripheral neural activity of CI users taking into account the non-linearity between stimulation current and loudness perception and the processes that may contribute to loudness beyond the peripheral system. The goal is to provide a new set of perceptual features that can be used by computational models for CIs, with the aim of narrowing the gap between simulated and real peripheral neural activity. To achieve our goal, a three-dimensional (3D) model of the cochlea was used to simulate this neural activity \citep{Nogueira2016, Alvarez2023, kipping2024, zhang2024}, and three experiments of loudness summation in real CI users were reproduced \citep{McKay2001, McKay2010, Zhou2012} to account for different aspects of the electrical stimulation.

\section{Computational loudness model}\label{sec_compmodel}

\begin{figure}[ht]
\centering
\includegraphics[width=1\textwidth]{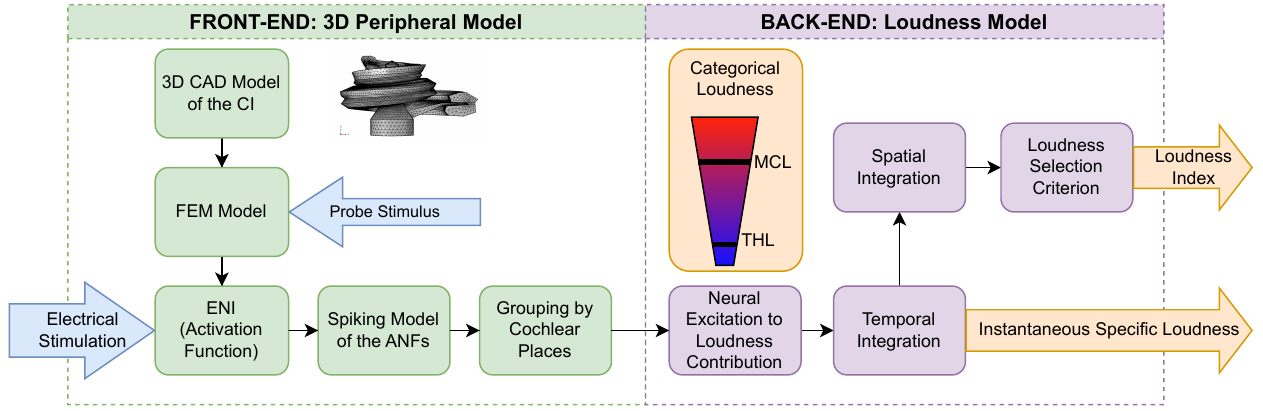}
\caption{Complete diagram of the computational loudness model. The front-end is the three-dimensional (3D) peripheral model of the cochlear implant (CI) constructed with a computer-aided design (CAD) software. A finite-element method (FEM) model was used to calculate the voltage distribution over a population of auditory nerve fibers (ANFs) given a probe stimulus of \(1~\text{µA}\) delivered through each electrode in monopolar mode. The electrode-nerve interface (ENI) was defined for the electrode array with the activation function, which was calculated deterministically using the voltage distribution and the morphology of the ANFs. The electrical stimulation was a train of biphasic pulses delivered through the electrode array. The neural response of the ANFs was obtained with a phenomenological spiking model coupled to the ENI and grouped in cochlear places, which are physiological representations of the auditory filters. The back-end is the proposed loudness model. The simulated peripheral neural activity is transformed into loudness contribution. A temporal integration model is used to obtain the instantaneous specific loudness, a proposed perceptual feature for computational frameworks using CIs. This feature is integrated across cochlear places to obtain the instantaneous loudness and the loudness index was defined as the 99\% percentile of it. The loudness index served to define a categorical loudness scale including the threshold of hearing (THL) and the most comfortable loudness (MCL) level.}
\label{LoudModel}
\end{figure}

As shown in Figure \ref{LoudModel}, the computational model consists of a 3D peripheral auditory model in the front-end, and the proposed loudness model in the back-end. The following sections offer a detailed explanation for each stage.

\subsection{Peripheral auditory model}

The 3D peripheral model consisted of: i) a deterministic ENI that included a finite-element method (FEM) model to obtain the excitation profile given electrical stimuli, ii) a phenomenological ANF model to simulate the spike activity produced by a stimulus in time, and iii) a final stage to represent the spike activity as peripheral neural excitation.

\subsubsection{Electrode-nerve interface}\label{subsubsec_ENI}

The 3D model of the implanted cochlea used in this work is largely based on the model presented in \cite{zhang2024} up to the FEM model. Only the ANF morphology was updated to the one described in \cite{Kalkman2022}. In this case, the myelinated segments in the peripheral and central processes, as well as the thinly myelinated soma, were considered completely insulated from external currents for technical reasons, leaving the nodes of Ranvier and the peripheral terminal as the available external stimulation places. We refer to these places as the stimulation compartments of the ANF.

The FEM model was used to obtain the voltage distribution \(V_0^E\) over a population of 30000 ANFs given a probe current \(I_0 = 1~\text{µA}\) delivered through each electrode \(E\). The electrode array virtually inserted in the 3D peripheral model was the slim modiolar (CI632) from the manufacturer Cochlear Ltd \textregistered. Electrodes were numbered from E1 (most basal) to E22 (most apical). The activation function for a probe current \(\text{AF}^E_{0,i,j}\) was calculated as shown in Eq. \ref{eqAF}, where \(i\) is the stimulation compartment index counting from the peripheral axis towards the central process, \(j\) is the ANF index, \(G\) is the axonal conductance between two adjacent stimulation compartments, and \(C\) is the stimulation compartment membrane capacitance.

\begin{equation}
\text{AF}_{0,i,j}^{E} = G_{i-1,i,j}\frac{V_{0,i-1,j}^{E}-V_{0,i,j}^{E}}{C_{i,j}} + G_{i,i+1,j}\frac{V_{0,i+1,j}^{E}-V_{0,i,j}^{E}}{C_{i,j}}
\label{eqAF}
\end{equation}

\begin{figure}[ht]
\centering
\includegraphics[width=0.67\textwidth]{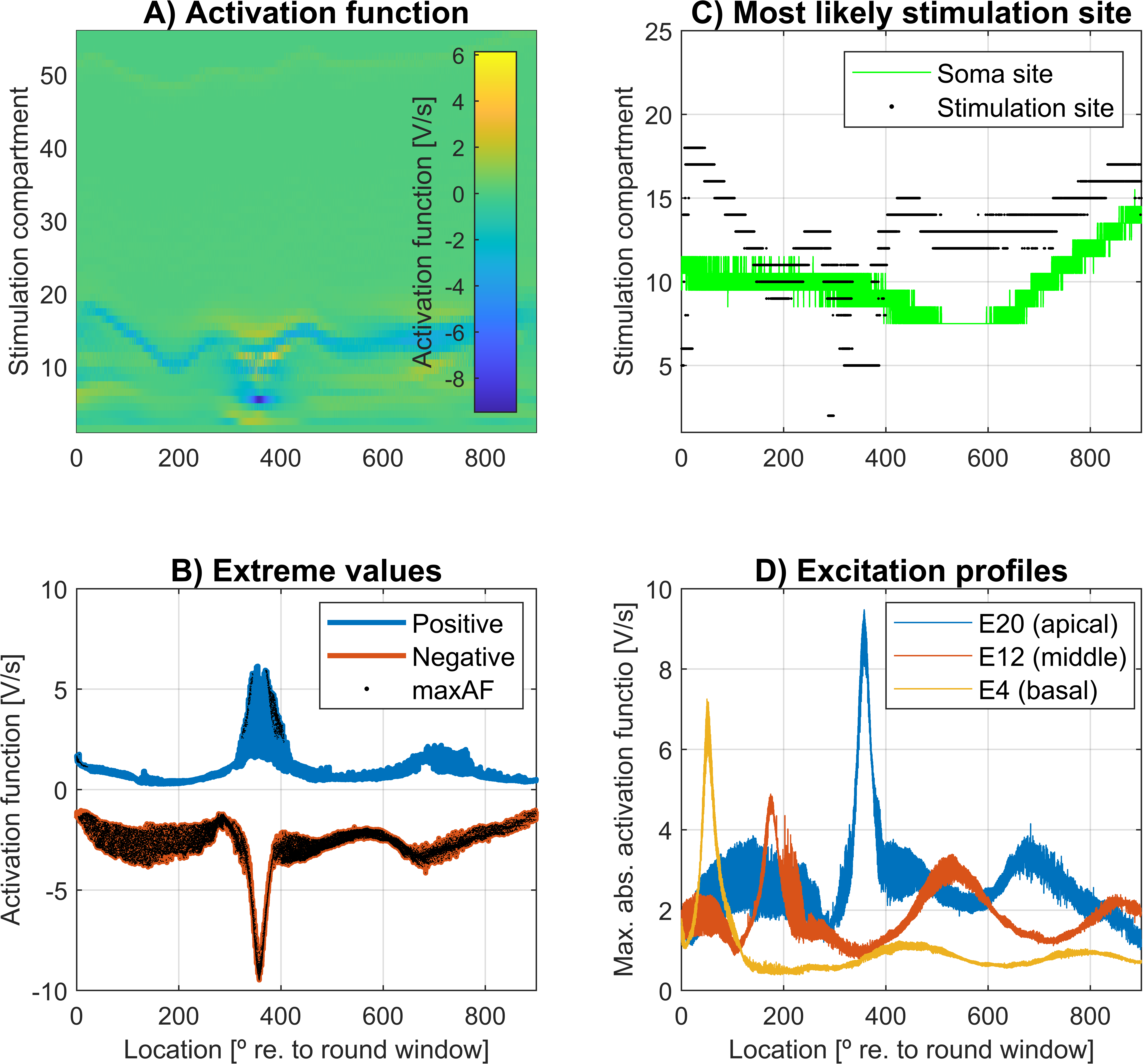}
\caption{Activation function obtained with the three-dimensional peripheral model using a probe current of 1 µA at electrode 20 (E20). A) Activation function \(\text{AF}_{0,i,j}\) in every stimulation compartment, \(i\), across the population of auditory nerve fibers (ANFs), \(j\). Stimulation compartments are numbered from most peripheral to most central. B) Most positive and most negative activation function values along each ANF, \(j\), with the selected values for the single-valued activation function \(\text{maxAF}_j\). C) Location of the most likely stimulation site. The stimulation compartments from where \(\text{maxAF}_j\) was selected. The location of the soma is indicated in green as a reference. D) Excitation profiles as the maximum absolute value of the activation function \(|\text{maxAF}_j|\), overlayed for electrodes E4, E12 and E20.}
\label{figAF}
\end{figure}

A special case was considered for the nodes of Ranvier adjacent to the soma where the axonal conductance \(G\) was manually adjusted to compensate for the abnormal high values of the activation function reported in \cite{kipping2024} and \cite{zhang2024}. Figure \ref{figAF}.A shows the activation function obtained for E20.

The activation function is directly related to the effectiveness of stimulation of an external current \citep{Rattay1999}. The advantage of using the activation function in an electrostatic field is that any proportional change \(\gamma^E={I^E}/{I_0}\) in the stimulation current, delivered through any electrode \(E\), results in a change of the same proportion in the voltage distribution and in the activation function. Therefore, the activation function resulting from a given stimulation current was calculated with Eq. \ref{eqTotalAF}.

\begin{equation}
\text{AF}_{i,j} = \frac{I^E}{I_0} \text{AF}^E_{0,i,j} = \gamma^E \cdot \text{AF}^E_{0,i,j}
\label{eqTotalAF}
\end{equation}

To reduce computational effort, the ``most likely stimulation site'' \(i_\text{max}\) was selected for each ANF by locating the stimulation compartment with the highest absolute value of the activation function \citep{Alvarez2023, kipping2024, zhang2024}. Therefore, \(\text{maxAF}_j = \text{AF}_{i_\text{max},j}\), where \(i_\text{max} = \text{argmax}_i(|\text{AF}_{i,j}|)\). Figure \ref{figAF}.B shows the most positive and most negative values of \(\text{AF}_{i,j}\) along the cochlea and highlights in black the values selected for \(\text{maxAF}_j\). Figure \ref{figAF}.C shows where the most likely stimulation site is located.

We defined the excitation profile as the magnitude of the single-valued activation function \(|\text{maxAF}_j|\) for each electrode \(E\). Figure \ref{figAF}.D shows the excitation profiles of a basal, middle and apical electrode obtained with the 3D peripheral model.

The external current \(I^E\) delivered through the CI632 is shown in current units (CUs) as used by the manufacturer. Use Eq \ref{EqCU2I} to obtain its value in Amperes.

\begin{equation}
I^E = 10^{-5} \cdot 175^{\frac{\text{CU}}{255}}
\label{EqCU2I}
\end{equation}

\subsubsection{Phenomenological auditory nerve fiber model}\label{subsubsec_ANF}

The phenomenological model used in this work was based on the model proposed by \cite{Joshi2017} including the randomization of thresholds and parametrization from \cite{Kipping2022} to account for a population model.

This phenomenological model consisted of two exponential integrate-and-fire circuits. Originally, both circuits were described as corresponding to the peripheral and central axons of a single neuron assuming that the peripheral axon is excited by the cathodic phase and inhibited by the anodic phase of a biphasic pulse, while the opposite is true for the central axon. However, this might be a consequence of the ENI geometry as observed in studies investigating the polarity effect of different types of pulses using 3D models \citep{Kalkman2022}. As shown in Figures \ref{figAF}.B and \ref{figAF}.C, the single-valued activation function \(\text{maxAF}_j\) took mainly negative values in both peripheral and central processes making them both excitable by the cathodic phase. Only around the main lobe, the positive values of \(\text{maxAF}_j\) corresponded to central processes that were excited by the anodic phase. For this reason, instead of using the \textit{peripheral axon} and \textit{central axon} denomination, both circuits will be referred to as \textit{cathodic-excitatory} and \textit{anodic-excitatory}, respectively, as in \cite{Alvarez2023}.

The membrane potential \(U(t)\) was computed every time step \(t\) as a function of the extracellular current \(I_j(t)\) as shown in Eq \ref{EqmembraneV}, where \(\hat{C}\) is the membrane capacitance, \(h(U)\) is a passive filter that includes the exponential upswing of the membrane voltage during a spike generation, \(I_a(t)\) represents the sub and suprathreshold adaptation currents, and \(I_N(t)\) is the Gaussian noise that introduces stochasticity \citep{Joshi2017, Kipping2022, Alvarez2023}.

\begin{equation}
\hat{C} \cdot \frac{dU(t)}{dt} = h(U) - I_a(t) + I_N(t) + I_j(t)
\label{EqmembraneV}
\end{equation}

Its value for the cathodic-excitatory circuit \(U^{\text{CE}}(t)\) and anodic-excitatory circuit \(U^{\text{AE}}(t)\) were calculated simultaneously, and independently, as described in \cite{Kipping2022}. The extracellular current was obtained with Eq \ref{EqIind}, where \(\alpha = 3.1826~\text{F}\) is a modeling factor that coupled the deterministic model of the ENI and the phenomenological model of the ANF population similar to \cite{Alvarez2023}. The value of the modeling factor was fitted to keep the stimulation current levels \(I^E\) in the range of real CI users (see section \ref{subsec_TMCL}).

\begin{equation}
I_j(t) = \alpha \cdot \text{maxAF}_j(t)
\label{EqIind}
\end{equation}

The single-valued activation function \(\text{maxAF}_j(t)\) was sampled at a rate of \(1~\text{MHz}\). The extracellular current \(I_j(t)\) for the cathodic-excitatory circuit \(I^{\text{CE}}_j(t)\) and the anodic-excitatory circuit \(I^{\text{AE}}_j(t)\) were calculated with Eq \ref{eqIndCircuits}, where \(\beta = 0.75\) is the inhibitory factor \citep{Joshi2017, Kipping2022, Alvarez2023}, \(I^+_j(t)\) contains only the positive values of \(I_j(t)\), and \(I^-_j(t)\) the negative values of \(I_j(t)\).

\begin{eqnarray}
\label{eqIndCircuits}
    \begin{cases}
    I^{\text{CE}}_j(t) = I^+_j(t) + \beta \cdot I^-_j(t) \\
    I^{\text{AE}}_j(t) = -(I^-_j(t) + \beta \cdot I^+_j(t)) 
     \end{cases}
\end{eqnarray}

Note that the signs in Eq \ref{eqIndCircuits} are swapped with respect to the original publication \citep{Joshi2017}. This way, the ANFs in the main lobe of the excitation profile would mainly be excited by the cathodic phase if the ``most likely stimulation site'' is located in the peripheral axon.

The phenomenological model returned the spiking times from each ANF, which was used to build the spike activity signal \(\text{SA}_j(t)\) with a bin size of \(200~\text{µs}\). Because the bin size was smaller than the lowest possible absolute refractory period of the phenomenologically modeled ANFs, this signal oscillated between ones and zeros depending on wether a spike was detected, or not, within each time bin.

\subsubsection{Peripheral neural excitation}\label{subsubsec_NE2}

The peripheral neural excitation is the representation of the spike activity across cochlear places \(x\). In this work, it was assumed that every ANF contributes the same to the overall loudness, therefore, the cochlear places were defined in groups of \(N = 750\) adjacent ANFs to obtain a total of 40 cochlear places. This amount approximates the number of independent auditory filters required to cover the human hearing spectrum \citep{Moore2003}.

\begin{figure}[ht]
\centering
\includegraphics[width=0.33\textwidth]{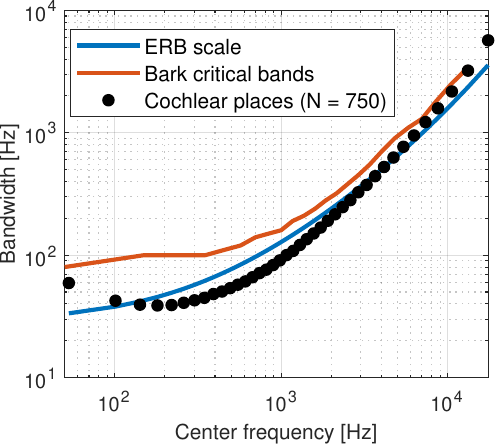}
\caption{Center frequencies and bandwidth of the 40 cochlear places defined in the three-dimensional peripheral model. The cochlear places are formed by groups of N = 750 adjacent auditory nerve fibers. Equivalent rectangular bandwidth (ERB) scale and the Bark critical bands are shown as a reference.}
\label{figCriticalbands}
\end{figure}

This physiological approach was compared to the auditory filters defined in psychoacoustics by estimating the center frequency and bandwidth of each cochlear place. Note that the 3D peripheral model does not use an homogeneous distribution of the ANF along the basilar membrane \citep{kipping2024, zhang2024}, instead, the ANF are arranged following the distribution shown in \cite{Spoendlin1989}. Additionally, the characteristic frequency of each ANF was estimated with the place-to-frequency equation from \cite{Greenwood1990}. As shown in Figure \ref{figCriticalbands}, this selection criterion of cochlear places fairly resembled the bandwidths of the Bark critical bands and the ERB scale.

With the cochlear places defined, the peripheral neural excitation \(E_x(t)\) was obtained with Eq \ref{EqExcProfile}.

\begin{equation}
E_x(t) = \frac{1}{N}\sum_{j \in x} \text{SA}_j(t)
\label{EqExcProfile}
\end{equation}

Because the spike activity is bounded by values between 0 and 1. The peripheral neural excitation can also be thought of as a signal that indicates the instantaneous firing probability in each cochlear place.

\subsection{Loudness model}\label{subsec_Loudmodel}

The proposed loudness model is based on the work of McKay and colleagues \citep{McKay2021} and consisted of four stages: i) the loudness contribution estimation, ii) the temporal integration model, iii) the cochlear place integration, and iv) the loudness index criterion.

\subsubsection{Loudness contribution estimation}\label{subsubsec_SLoud}

The relation between the peripheral neural excitation \(E_x(t)\) and the instantaneous loudness contribution \(\lambda_x(t)\) was modeled using Eq \ref{EqISL},  where \(k = 0.525\) is a linear scaling factor representing the relation between neural activity and loudness when the stimulation current is low, and \(\epsilon(t)\) represents the exponential increase at higher stimulation currents.

\begin{equation}
\lambda_x(t) = k \cdot \textit{E}_x(t) \cdot (1 + \epsilon(t))
\label{EqISL}
\end{equation}

The scaling factor \(k\) was fitted to obtain more intuitive values for the categorical loudness at the output of the loudness model. Please refer to section \ref{subsec_TMCL}.

This work assumes that the exponential increase in loudness happens at higher stages of the auditory pathway, therefore, the exponential factor \(\epsilon(t)\) was computed using Eq \ref{EqExpfactor}, where the overall neural activity was obtained as \(E_T(t) = \sum_xE_x(t)\), and the parameters \(E_0 = 12\) and \(m = 3.2\) were fitted to reproduce the logarithmic loudness growth slope presented in \cite{McKay2003}. This factor was close to 1 when the overall neural activity \(E_T(t)\) was focused on few cochlear places and exponentially increased as more cochlear places were recruited with higher stimulation currents.

\begin{equation}
\epsilon(t) = e^ \frac{E_T(t) - E_0}{m}
\label{EqExpfactor}
\end{equation}

The instantaneous loudness contribution provides a measure of how the peripheral neural activity contributes to loudness perception and was sampled every \(200~\text{µs}\).

\subsubsection{Temporal integration model}\label{subsubsec_LC}

The temporal integration model consists of the convolution between the instantaneous loudness contribution \(\lambda_x(t)\) and the asymmetric temporal integration window \(W(t)\). The convolution is performed for each cochlear places to obtain the instantaneous specific loudness \(\Lambda_x(t)\) as shown in Eq \ref{EqLt}.

\begin{equation}
\Lambda_x(t) = \lambda_x(t) * W(t)
\label{EqLt}
\end{equation}

The integration window was fitted as for normal hearing subjects \citep{Oxenham1994, Oxenham2001} as shown in Eq \ref{EqTIM}, where \(T_{b1} = 4.6~\text{ms}\) and \(T_a = 3.5~\text{ms}\) are time constants to account for temporal resolution, \(T_{b2}=16.6~\text{ms}\) is a time constant to account for forward masking effects, and \(w = 0.17\) is the weight between the these two effects.

\begin{eqnarray}
\label{EqTIM}
    \begin{cases}
    W(t) = (1 - w) \cdot e^{\frac{t}{T_{b1}}} + w \cdot e^{\frac{t}{T_{b2}}}, \quad  t < 0 \\
    W(t) = e^{\frac{-t}{T_a}}, \quad \quad \quad \quad \quad \quad \quad \quad  \quad ~~~ t \geq 0
    \end{cases}
\end{eqnarray}

This approach has also been used in loudness models for CI models \citep{McKay1998, McKay2013a, McKay2021}, with the argument that temporal integration occurs at a stage posterior to the peripheral auditory system.

\begin{figure}[ht]
\centering
\includegraphics[width=0.99\textwidth]{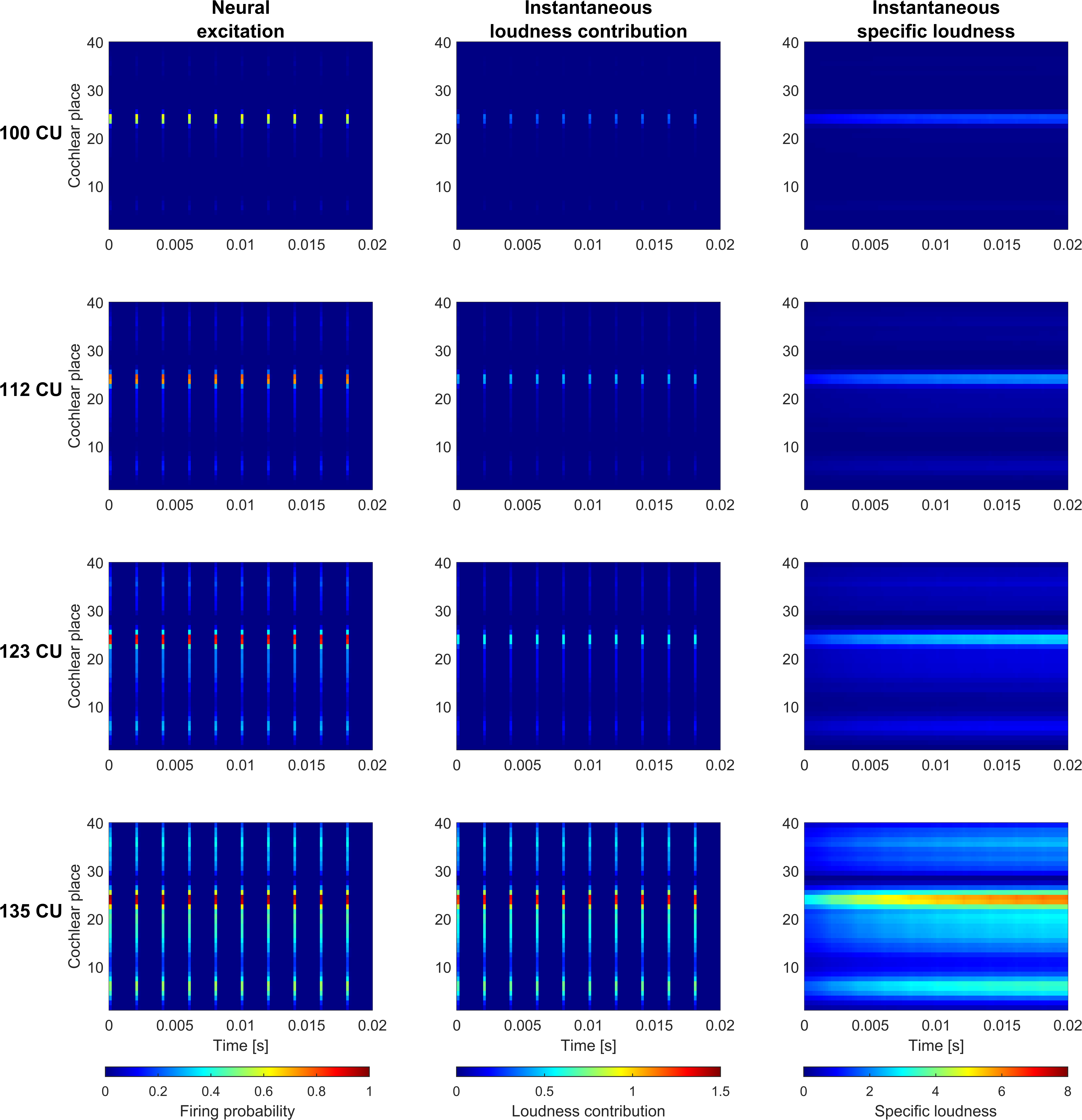}
\caption{Visual representation of the first \(20~\text{ms}\) of the signal throughout the loudness recruitment model given a pulse train stimulus of biphasic pulses delivered through the electrode 20 at a rate of \(500~pps\). The left column shows the neural excitation as a function of the firing efficiency across cochlear places. The middle column shows the instantaneous loudness contribution calculated from the neural excitation. The right column shows the instantaneous specific loudness after applying the temporal integration. The rows are organized from top to bottom representing the stimulation in current units (CUs).}
\label{figInstSpecL}
\end{figure}

Figure \ref{figInstSpecL} provides a visual representation of the neural excitation (left column), instantaneous loudness contribution (center column), and instantaneous specific loudness (right column), after delivering a pulse train stimulus at a rate of \(500~\text{pps}\) through E20. The rows are organized top to bottom from lower to higher stimulation currents. Notice that the instantaneous loudness contribution and instantaneous specific loudness dramatically increased between the rows corresponding to \(123~\text{CU}\) and \(135~\text{CU}\) because of the exponential factor that became relevant after more cochlear places presented neural activity (see Eq \ref{EqISL} and \ref{EqExpfactor}).

\subsubsection{Cochlear place integration}\label{subsubsec_CPI}

Similar to loudness models based on critical bands \citep{Zwicker1957, Zwicker1980, moore1997, Glasberg2002, iso2017532}, the instantaneous loudness \(l(t)\) has been calculated by summing up the individual contributions to loudness across the cochlear places as shown in Eq \ref{EqIL}.

\begin{equation}
l(t) = \sum_x \Lambda_x(t).
\label{EqIL}
\end{equation}

\subsubsection{Loudness index criterion}\label{subsubsec_LIndex}

It was assumed that the instantaneous loudness \(l(t)\) provided information of loudness perception changes over time. However, in order to compare various stimuli in terms of a categorical loudness scale, a single value had to be derived from the instantaneous loudness. The criterion to obtain this value, defined as loudness index \(L\), was the 99th percentile of the instantaneous loudness across the whole stimulus \citep{Rennies2010, Francart2014}. This criterion works for short and invariant stimuli.

\begin{figure}[ht]
\centering
\includegraphics[width=0.33\textwidth]{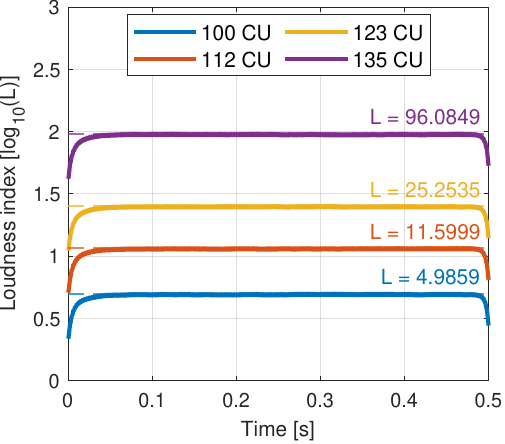}
\caption{Instantaneous loudness and loudness index (L) obtained with a pulse train stimulus of biphasic pulses delivered through the electrode 20 (E20) at a rate of \(500~\text{pps}\). The stimulation is represented in current units (CUs). The y-axis represents the loudness index in a logarithmic scale.}
\label{FigInstaL}
\end{figure}

Figure \ref{FigInstaL} shows the instantaneous loudness obtained after the cochlear place integration of the signals shown in Figure \ref{figInstSpecL}. It also shows the loudness index \(L\) obtained for each case. A stimulus with \(L=100\) is considered twice as loud as one with \(L=50\). The amount of peripheral neural activity might be better related to the logarithm of the loudness index. Notice that the logarithmical index increases linearly with the stimulation current in CUs at lower stimulation levels, between \(100~\text{CU}\) and \(123~\text{CU}\). This linear increase is synonym to the power function between stimulation current and loudness perception described previously in the introduction. At higher stimulation levels, between \(123~\text{CU}\) and \(135~\text{CU}\), a faster increase shows the exponential relation defined with Eq \ref{EqISL}.

\subsection{Categorical loudness}\label{subsec_TMCL}

\begin{figure}[ht]
\centering
\includegraphics[width=0.67\textwidth]{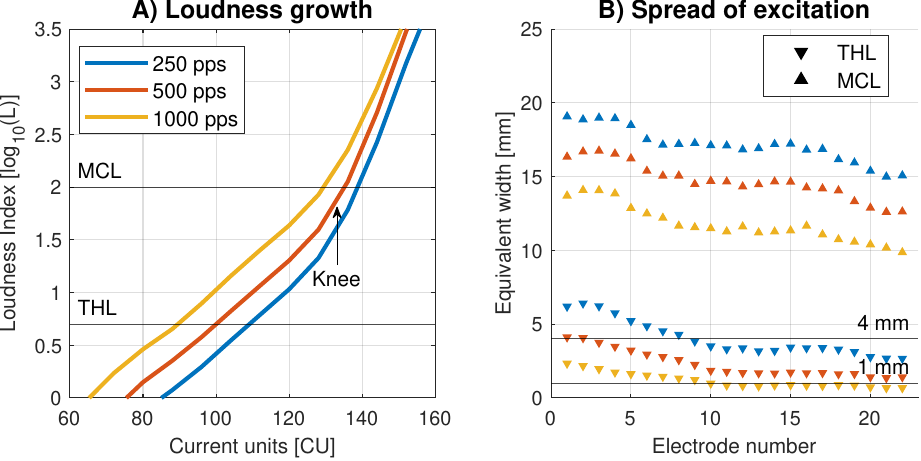}
\caption{Definition of threshold of hearing (THL) and most comfortable loudness (MCL) levels for the computational model. A) shows the loudness growth function (LGF) for electrode 20 at 250, 500 and 1000 pps. The y-axis shows the loudness index (L) in a logarithmic scale and the x-axis the stimulation index in current units (CUs). THL and MCL levels were chosen with the proposed new method. Notice the knee point of the LGF at 500 pps happening right before reaching the MCL level. B) shows the equivalent excitation width along the basilar membrane when stimulating at THL and MCL levels. The equivalent width is the cumulative area where the auditory nerve fibers are excited with a firing efficiency equal or higher than 50\%. The 1 mm and 4 mm are depicted as a reference to previous criteria for selecting THL and MCL levels \citep{Briaire2006, SnelBongers2013}.}
\label{figTHLMCL}
\end{figure}

To define a categorical loudness scale, the LGF was obtained for each electrode in a wide range of current levels using a train of biphasic pulses at a rate of 250, 500 and 1000 pps. Experiments with CI users showed that the current compensation needed for equal loudness perception between two stimuli that only differ in the stimulation rate is overall higher at THL than at MCL levels \citep{Shannon1985,McKay1998, McKay2001, McKay2003}. Additionally, when stimulating at a rate of 500 pps, it was measured that the knee point between the linear and the exponential increase in the loudness growth curve was located close and below the MCL level \citep{McKay2003, McKay2010}. As shown in Figure \ref{figTHLMCL}, the knee point was located in the LGF curve corresponding to 500 pps. Then, the scaling factor \(k\), in Eq \ref{EqISL}, was adjusted to locate the \(L=100\) mark above it. This adjustment did not affect the shape of the LGF, instead, offered a more intuitive index at the output of the loudness model similar to \cite{McKay2010}, where the THL and MCL levels corresponded to loudness indices of 5 and 100, respectively.

In other studies, MCL and THL levels were chosen regarding the spread of excitation produced by a pulse \citep{Briaire2006, SnelBongers2013, Kalkman2014, Alvarez2023, kipping2024, zhang2024}. The current at which this pulse excited an equivalent width of \(4~\text{mm}\) of the basilar membrane with a firing efficiency of 50\% or higher was considered the MCL level. The THL level was either set with a similar procedure but using an equivalent width of \(1~\text{mm}\), or by defining a fraction of the MCL level. Figure \ref{figTHLMCL}.B shows the excitation spread obtained with the proposed computational model. Notice that THL levels using this criterion corresponded to the \(1~\text{mm}\) equivalent excitation width for middle and apical electrodes (\(E \geq 8\)) as suggested by \cite{SnelBongers2013}, however, the equivalent excitation width at MCL is much higher than the \(4~\text{mm}\) mark.

\begin{figure}[ht]
\centering
\includegraphics[width=0.67\textwidth]{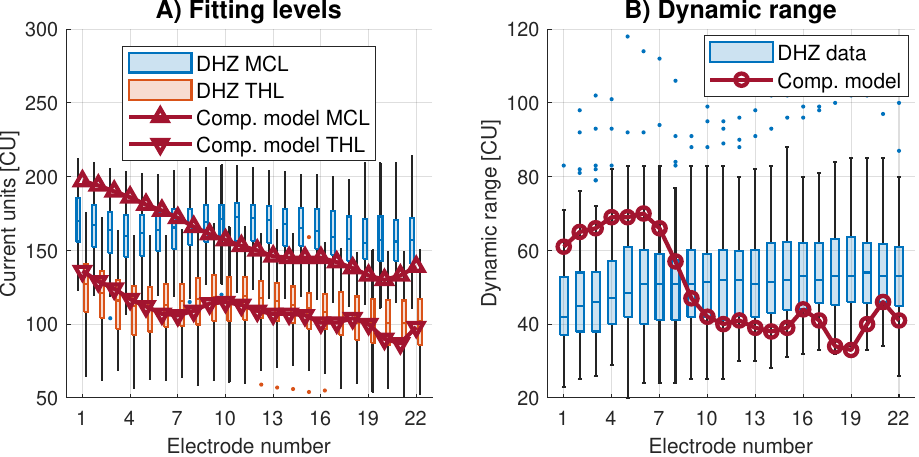}
\caption{Fitting of the computational model with a stimulation rate of 1000 pps compared to data of CI632 users taken from the \textit{Deutches Hörzentrum Hannover} (DHZ). Only the first fitting after the first year of implantation was taken into account for each individual. The DHZ data also combines different settings of stimulation rates, pulse phase duration and inter-phase gap. The x-axis represent the electrode number from most basal (E1) to most apical (E22). A) Comparison of the threshold of hearing (THL) and most comfortable loudness (MCL) levels obtained for the computational model and the data from DHZ in current units (CUs). B) Comparison of the dynamic range, as the MCL and THL level difference in CUs, between the computational model and DHZ data. Fitting of the computational model is shown with dark red lines and markers (triangles and circles). The boxplots were used to represent the DHZ data, which indicate the median with the central mark, and the 25th and 75th percentiles with the bottom and top edges of the box, respectively. The whiskers extend to the most extreme data points not considered outliers, and the outliers are plotted individually using the dots.}
\label{figFit}
\end{figure}

Additionally, the modeling factor \(\alpha\) in Eq \ref{EqIind} was adjusted to locate the THL levels when stimulating at 1000 pps in the range of real CI users. The reference data was collected from the \textit{Deutches Hörzentrum Hannover} (DHZ) databank of subjects implanted with the CI632. Figure \ref{figFit}.A shows THL and MCL levels obtained with the computational model on top of the THL and MCL levels distribution from the DHZ. \ref{figFit}.B shows the dynamic range in CUs as the difference between MCL and THL levels. This reference data is not limited to stimulation rates of 1000 pps or any particular phase duration of the biphasic pulse, however, it served the purpose to work with more realistic stimulation current levels.

\section{Experiments}\label{secExperiments}

The experiments performed to evaluate the loudness model used only sequential stimulation of \(500~\text{ms}\) long stimuli. The electrical stimulus consisted of cathodic leading biphasic pulses in monopolar mode. The phase duration was set to \(25~\text{µs}\) with an inter-phase gap of \(10~\text{µs}\).

The experiments were chosen to evaluate the effects of rate, electrode separation and amplitude modulation in the overall loudness perception. Then, the performance was compared to the performance of real CI users \citep{McKay2001, McKay2010, Zhou2012}.

\subsection{Experiment 1: Effect of rate of stimulation}\label{subsec_ERS}

In CI users, it was observed that slopes of the stimulation current versus stimulation rate at THL levels are steeper than those at MCL levels \citep{McKay1998, Zhou2012}. This leads to an increase in the dynamic range with higher stimulation rates. 

In this work, an experiment performed by \cite{Zhou2012} was reproduced. First, THL and MCL levels were calculated for each electrode in steps of \(1~\text{CU}\) using pulse trains at rates of 156, 313, 625, 1250, 2500, 5000 pps. Then, each value was normalized relative to the results at 156 pps. Since \cite{Zhou2012} did not find any significant effect of the electrode position, the electrodes of the computational model were not classified as basal, middle or apical.

\subsection{Experiment 2: Effect of electrode separation}\label{subsec_EES}

The practical method proposed by \cite{McKay2003} suggests that the loudness contribution of biphasic pulses is independent of the relative position of the stimulating electrodes along the electrode array. It is argued that it is the result of two counteracting effects. Assuming sequential stimulation, if two electrodes have an overlapping excitation pattern, the refractory effect of the ANFs would reduce the neural activity of subsequent pulses in the overlapping cochlear places. However, when the temporal integration occurs, these overlapping cochlear places would have a more frequent spiking activity which increases the loudness contribution \citep{McKay2001, McKay2003, McKay2021}. To validate this feature with the proposed computational model, an experiment performed in CI subjects by \cite{McKay2001} was reproduced. In this experiment, the stimulation current reduction in dB needed to equalize the loudness perception between a two-electrode interleaved stimulus and a single-electrode stimulus was calculated. Note that \cite{McKay2001} performed this experiment with CI users implanted with the CI24M from Cochlear Ltd \textregistered. The electrodes in the CI24M array are separated by \(0.75 \text{mm}\) while the electrodes in the CI632 are separated by \(0.69~\text{mm}\). However, this aspect should not be relevant in the context of this experiment. 

The experiment was set up using stimulation rates of 500 and 1000 pps and was repeated using each electrode of the computational model as the reference. For each reference electrode, the stimulation current was adjusted in steps of \(1~\text{CU}\) to reach MCL and THL levels. Then, the loudness index at half of their dynamic range in CUs was calculated to define a medium loudness (MDL) level. Three additional electrodes for each reference electrode were chosen with a separation of 1 electrode, 3 electrodes and 10 electrodes in the array. For E1 to E11 the additional electrodes were selected towards the apex while for E12 to E22 towards the base. MCL and MDL levels were obtained for the additional electrodes as well.

A single period of the two-electrode interleaved stimulus consisted of a single pulse in the reference electrode followed by another single pulse in the additional electrode with an onset delay of \(400~\text{µs}\). This pair of pulses were initially presented at the stimulation current that elicited the target level (MCL or MDL) in each electrode and were presented at the required rate (500 or 1000 pps) for the total duration of the stimulus. Then, the stimulation currents in both electrodes were simultaneously reduced in steps of \(1~\text{CU}\) until the loudness index reached the target level. This reduction represented the loudness summation of adding the second pulse in the two-electrode interleaved stimulus and can be measured in dB or percentage of the dynamic range.

\subsection{Experiment 3: Effect of amplitude modulation}\label{subsec_EAM}

In amplitude modulation detection experiments with CI users, the loudness balancing between modulated and non-modulated pulse trains depends on the carrier rate, the stimulation level and the modulation depth. \cite{McKay2010} investigated these effects and observed that the perceived loudness of modulated stimuli is affected mainly by the pulses around the peak level, especially at higher stimulation levels, higher modulation depths and lower carrier rates. This might be caused by the expansive nature of the neural excitation to loudness contribution function (loudness grows faster with more neural activity) \citep{McKay2001}. The implications of these observations is that loudness cues might be used by participants in amplitude modulation and amplitude modulation rate detection experiments if they are not handled properly \citep{McKay2010, Chatterjee2011}.

This experiment is a reproduction of the loudness balancing experiment performed by \cite{McKay2010}. The amplitude modulated stimuli consisted of a train of biphasic pulses presented at a carrier rate of 500, 1000 or 8000 pps. The amplitude of the pulses was modulated by a sinusoidal signal at \(250~\text{Hz}\) when the carrier rate was \(500~\text{pps}\), and \(500~\text{Hz}\) when the carrier rate was 1000 or 8000 pps. The phase of the modulation signal was always fitted to get the maximum amplitude in the first pulse of every modulated stimulus. The modulation depth, defined as the difference in current between the highest (peak) and the lowest (valley) amplitude, was varied between 5, 10, 15 and 20 CU. The mean amplitude would be the middle point between the peak and the valley in CU.

First, the non-modulated stimuli was presented at each carrier rate on each reference electrode to find the MCL and THL levels. Then, the stimulation currents that represented 90\% and 60\% of the dynamic range were chosen as references in steps of \(1~\text{CU}\) and the loudness elicited was noted for each case. In the next steps, the modulated stimuli were constructed using the before mentioned modulation depth and setting the peak current to both reference stimulation currents. In total, there were 24 different modulated stimuli for each electrode combining the three carrier rates (500, 1000 and 8000 pps), two current references (60\% and 90\% of the dynamic range) and four modulation depths (5, 10, 15 and 20 CU). Finally, each modulated stimulus was varied in steps of \(1~\text{CU}\) until it reached the loudness of the respective non-modulated signal with the same carrier rate and dynamic range level.

\section{Results}\label{secResults}

\subsection{Experiment 1: Effect of rate of stimulation}\label{subsec_ResERS}

\begin{figure}[ht]
\centering
\includegraphics[width=0.67\textwidth]{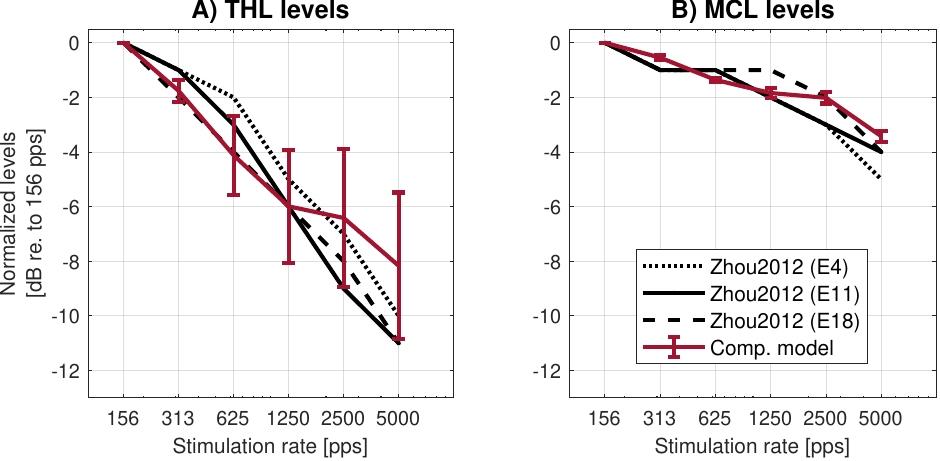}
\caption{Simulated results for the effect of rate of stimulation in cochlear implant (CI) users. A) Threshold of hearing (THL). B) Most comfortable loudness (MCL). THL and MCL levels are normalized to the results at 156 pps. Results with the computational loudness model are shown in red with the error bars representing the standard deviation across the electrodes. Values shown in black were taken from data of real CI users' performance. Mean values were reprinted from Figure 2 in \cite{Zhou2012}, where the basal electrodes (E4) are shown with the dotted line, middle electrodes (E11) with the solid line and apical electrodes (E18) with the dashed line.}
\label{figResultsE1}
\end{figure}

The normalized THL and MCL values were plotted against the stimulation rate in Figure \ref{figResultsE1}. It shows the results obtained with the computational model for the THL and MCL levels with the dark red lines. The line crosses the mean values across electrodes while the standard deviation is shown with the error bars. Reference mean values were visually approximated from \cite{Zhou2012} and reprinted in black lines. Note that \cite{Zhou2012} originally differentiate the results regarding basal (E4), middle (E11), and apical (E18) electrodes and they are shown with the dotted, solid and dashed line, respectively.

Additionally, the results were grouped into a below, and above, \(1000~\text{pps}\) of stimulation rate, respectively. A line was fitted to the THL and MCL in both groups to obtain the level versus stimulation rate slope as described in \cite{Zhou2012}. Table \ref{tab_Exp1} shows the slope change as the difference between the above \(1000~\text{pps}\) (\(>1000\)) and the below \(1000~\text{pps}\) (\(<1000\)) group.

\begin{table}[ht]
\caption{Normalized level versus stimulation rate slope obtained across electrodes with the computational loudness model (Comp. model). Results were separated by stimulation rates below 1000 pps (\(<1000\)) and above 1000 pps (\(>1000\)) at threshold of hearing (THL) and most comfortable loudness (MCL) levels. The slopes represent how much the current has to be reduced to reach the target level (THL or MCL) when doubling the stimulation rate. Difference between ``\(<1000\)'' and ``\(>1000\)'' shows how much this slope increases at high stimulation rates.}\label{tab_Exp1}
\begin{tabular*}{\textwidth}{@{\extracolsep\fill}lcccccc}
\toprule%
& \multicolumn{3}{@{}c@{}}{Slopes at THL levels [dB/doubling]} & \multicolumn{3}{@{}c@{}}{Slopes at MCL levels [dB/doubling]} \\\cmidrule{2-4}\cmidrule{5-7}%
From: & $<1000$ & $>1000$ & Difference & $<1000$ & $>1000$ & Difference \\
\midrule
Comp. model  & 2.06 & 1.09 & -0.97 & 0.68 & 0.79 & 0.11\\
Zhou2012 \footnotemark[1] & 1.50 & 2.66 & 1.16 & 0.50 & 1.28 & 0.78\\
\botrule
\end{tabular*}
\footnotetext[1]{Data taken and reinterpreted from \cite{Zhou2012}.}
\end{table}

\begin{figure}[ht]
\centering
\includegraphics[width=0.33\textwidth]{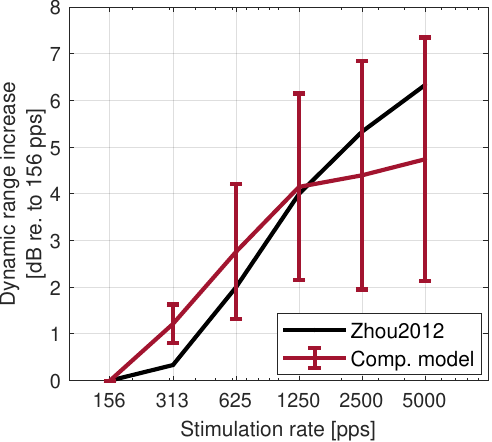}
\caption{Dynamic range increase when doubling the stimulation rate relative to \(156~\text{pps}\). The red line represent the results from the computational loudness model with the error bars showing the standard deviation across electrodes. The black line represent the dynamic range increase estimated from the results published by \cite{Zhou2012}.}
\label{figResultsE1DR}
\end{figure}

The dynamic range increase across electrodes when using the proposed computational model is shown in Figure \ref{figResultsE1DR} with the red line including the error bars representing the standard deviation across electrodes. The black line shows the dynamic range increase derived from \cite{Zhou2012} as a reference.

\subsection{Experiment 2: Effect of electrode separation}\label{subsec_ResEES}

\begin{figure}[ht]
\centering
\includegraphics[width=0.67\textwidth]{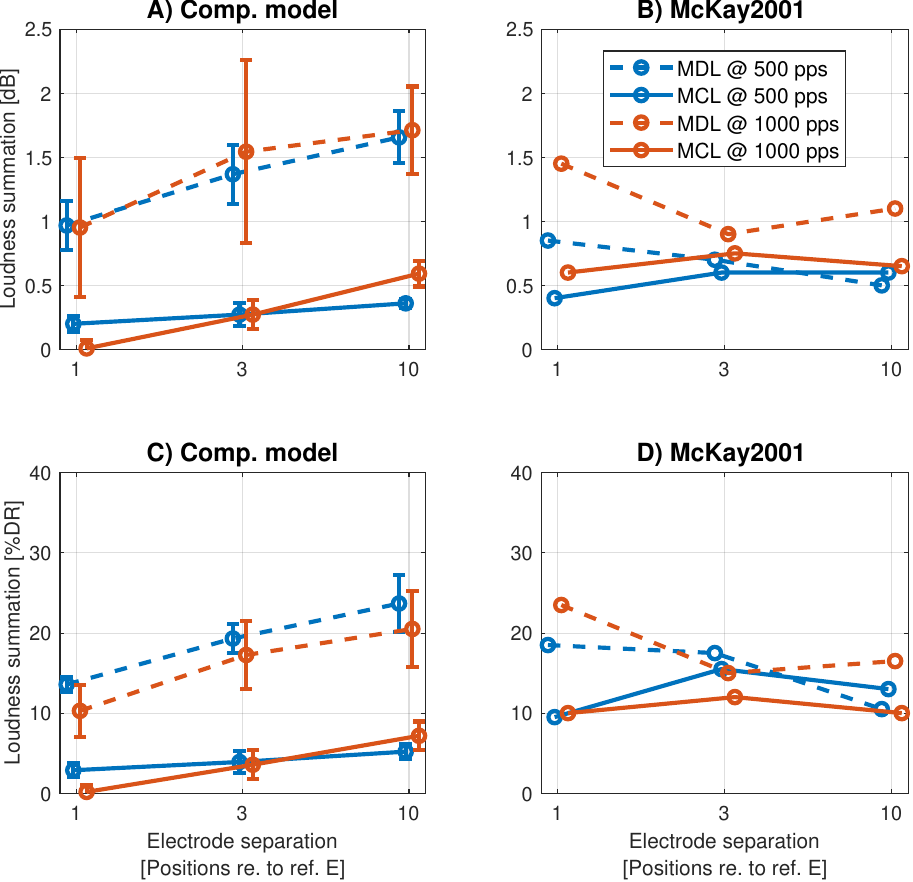}
\caption{Loudness summation with a secondary pulse in a two-electrode interleaved stimulus. The x-axis indicates the electrode separation between the reference and the additional electrode. Top panels (A and B) show the loudness summation in dB relative to the level of the reference signal (a single electrode pulse train). Bottom panels (C and D) show the loudness summation as a percentage of the dynamic range (\%DR). Left panels (A and C) show the results with the proposed computational model with the error bars representing the standard deviation across electrodes. Right panels (B and D) show the reprinted values from Figures 1 and 2 in \cite{McKay2001}. Middle loudness (MDL; at 50\% of the DR) and most comfortable loudness (MCL) levels are printed with dashed and solid lines, respectively. Blue and red colors correspond to a stimulation rates of \(500~\text{pps}\) and \(1000~\text{pps}\), respectively.}
\label{figResultsE2}
\end{figure}

The results from the computational loudness model are shown in Figures \ref{figResultsE2}.A and \ref{figResultsE2}.C. The symbols (triangles for MCL and circles for MDL) represent the mean loudness summation across electrodes and the error bars represent the standard deviation. Solid and dashed lines show the results with \(500~\text{pps}\) and \(1000~\text{pps}\), respectively. The values in Figures \ref{figResultsE2}.B and \ref{figResultsE2}.D were reprinted from Figures 1 and 2 in \cite{McKay2001}, respectively, of the CI24M users that were tested using the monopolar mode. Top figures (\ref{figResultsE2}.A, \ref{figResultsE2}.B) show the loudness summation in dB while the bottom figures (\ref{figResultsE2}.C, \ref{figResultsE2}.D) show it as a percentage of the dynamic range.

\subsection{Experiment 3: Effect of amplitude modulation}\label{subsec_ResEAM}

\begin{figure}[ht]
\centering
\includegraphics[width=0.67\textwidth]{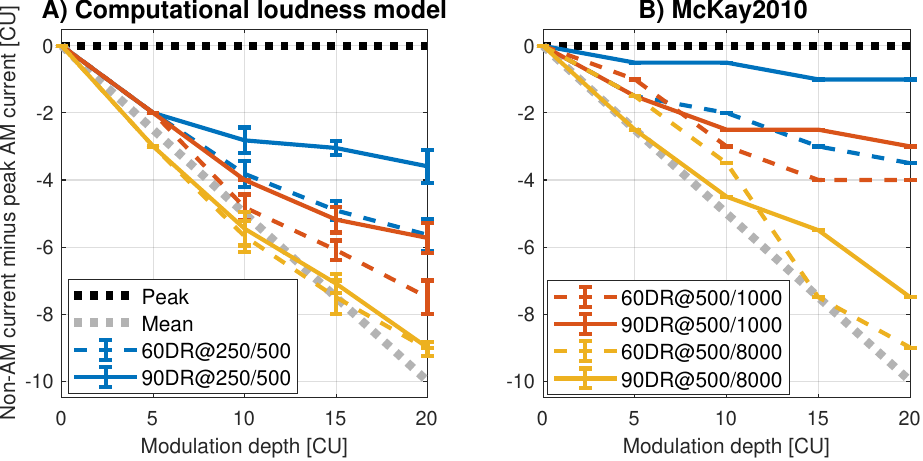}
\caption{Difference measured in current units (CU) between a non-modulated stimulus and a modulated stimulus to reach equal loudness. The x-axis indicates the modulation depth in CU and the colors represent the different carrier rates (blue for \(500~\text{pps}\), red for \(1000~\text{pps}\) and yellow for \(8000~\text{pps}\). In panel A, the lines show the results with the computational loudness model with the error bars representing the standard deviation across electrodes. Panel B is the reprinted data from Figure 2 in \cite{McKay2010}. Solid and dashed lines are used for results at 90\% of the dynamic range (90DR) and 60\% of the dynamic range (60DR), respectively. The modulation frequency used was equal to \(250~\text{Hz}\) when the carrier rate was \(500~\text{pps}\) (250/500), and \(500~\text{Hz}\) when the carrier rate was \(1000~\text{pps}\) (500/1000) or \(8000~\text{pps}\) (500/8000). Black dotted lines and gray dotted lines show the values as if the peak current, or mean current, of the modulated stimulus was equal to the non-modulated stimulus current.}
\label{figResultsE3}
\end{figure}

Figure \ref{figResultsE3} shows the mean differences in peak stimulation level needed to obtain the same loudness between a non-modulated pulse train and a modulated pulse train across electrodes in comparison to data reported by \cite{McKay2010}. Triangles are used for 90\% of the dynamic range and circles for 60\% of the dynamic range. The error bars represent the standard deviation. Blue, red and yellow colors represent the carrier rate of 500, 1000 and 8000 pps, respectively. There are two additional dotted lines that show the difference when the modulated stimulus would have a peak current (black), or mean current (gray), equal to the non-modulated stimulus.Figure \ref{figResultsE3}.A shows the result obtained with the computational loudness model and the Figure \ref{figResultsE3}.B shows the result reprinted from Figure 2 in \cite{McKay2010}.

\section{Discussion}\label{disc}

The most important contribution of the computational loudness model presented in this work is that it offers a tool to estimate the categorical loudness from the simulated peripheral neural activity. In contrast, other state-of-the-art models compute the loudness directly from the electrical pulses of the CI, without explicitly considering a realistic ENI \citep{Francart2014, Langner2020, McKay2021}. Including a model of the peripheral system allows to account for more physiological parameters related to the implantation such as the electrode insertion, neural health and cochlear geometry.

In order to perform simulations with computational models that are closer to what is really happening in the human auditory system, it is important to calibrate the simulated neural activity to specific categorical loudness levels such as THL and MCL. This work highlights that the commonly used \(4~\text{mm}\) criterion proposed by \cite{Briaire2006} underestimates the area of excitation, or amount of neural activity, needed to reach MCL levels. At MCL levels, many regions of the cochlea have to be stimulated at the same time, and when these levels are reached with a single electrode, it means that the stimulation profile becomes much broader than \(4~\text{mm}\) (see Figures \ref{figInstSpecL} and \ref{figTHLMCL}.B). This is also observed with more complex stimuli where the current delivered through the electrodes at hearing ranges may produce an excitation spread larger than the spacing of the electrodes in the array \citep{Friesen2001}. 

\subsection{Performance regarding rate of stimulation, electrode separation and amplitude modulation}

In general, the predictions of the computational model in three different loudness experiments corresponded qualitatively with data from real CI users' performance. These experiments investigated the effects in loudness regarding the rate of stimulation, electrode separation and amplitude modulation.

In experiment 1, the model predictions for the normalized MCL levels as a function of stimulation rate matched the data of \cite{Zhou2012} very accurately (see Figure \ref{figResultsE1}.B) and corresponded well with the data for normalized THL levels at stimulation rates below 1000 pps (see Figure \ref{figResultsE1}.A). The model also agreed with data in the monotonic increase of the dynamic range when incrementing the stimulation rate (see Figure \ref{figResultsE1DR}). The discrepancies in THL levels at higher stimulation rates, together with the normalized level versus stimulation rate slope (see Table \ref{tab_Exp1}), could be traced back to limitations in the peripheral model and are discussed in the following subsection.

Results from experiment 2 showed minimal effect of the electrode separation in loudness summation with the two-electrode interleaved stimulus. Despite the curves being tilted suggesting a loudness summation increase with distance, the effect was always below 1~dB. This observation agrees with data from \cite{McKay2001} and also aligns with the hypothesis that the loudness contribution of the biphasic pulses in a two-electrode interleaved stimulus does not depend on the electrodes' relative position. Note that this does not explain loudness summation when the biphasic pulses are presented simultaneously, for example, when using current steering \citep{Langner2020}. This particular scenario was out of the scope of the current study, however, it can be accounted by extending Eq \ref{eqTotalAF} as the sum of the electrical contributions from all the electrodes as follows: \(\text{AF}_{i,j} = \sum_E \gamma^E \cdot \text{AF}^E_{0,i,j}\). This is possible because loudness is being predicted from the peripheral neural activity and it was assumed that the ENI behaves as in an electrostatic field. Quantitatively, the effect of stimulation level in loudness summation was better differentiated with the loudness model. This is expected because, at higher stimulation levels, changes in the current produce larger changes in the loudness perception due to the exponential factor (see Eq \ref{EqISL}). It is possible to reduce this effect by adjusting the parameters \(m\) and \(E_0\) in Eq \ref{EqExpfactor}.

Regarding experiment 3 on amplitude modulation, results with the computational model also align with the observations made by \cite{McKay2010}. The difference between the non-modulated current and the modulated peak current was generally lower with higher carrier rates and higher stimulation currents. This indicates that the loudness index, and loudness perception in case of real CI users, was mainly affected by the pulses with higher stimulation current. The model agreed with literature in that loudness balancing between modulated and non-modulated trains of pulses is needed to avoid loudness cues \citep{McKay2010, Chatterjee2011}. Quantitatively, the computational model tends to predict higher current differences between modulated and non-modulated stimuli than reported for CI users (see Figure \ref{figResultsE3}.A and \ref{figResultsE3}.B). This indicates that the MCL level could be selected further up from the knee point indicated in Figure \ref{figTHLMCL}. This would have made the exponential factor more relevant in stimuli at 60\% and 90\% of the dynamic range, therefore, smaller current changes would have produced larger changes in the categorical loudness.

As a final remark, quantitative differences between data and results in experiments 2 and 3 could be reduced with a ``more accurate'' selection of the THL and MCL levels. However, given the intersubject variability in CI users, qualitative results were the priority in this study. 

\subsection{Limitations of the peripheral model}

Having a combination of a deterministic and a phenomenological model of the peripheral auditory system saves plenty of computational resources compared to a fully deterministic model \citep{Rattay2001, Potrusil2020, Kalkman2022, Brochier2022}. Additionally, many phenomenological approaches also account for the stochastic behavior of the peripheral neural activity, which is an advantage over deterministic models \citep{Alvarez2023, kipping2024, zhang2024, deNobel2024}. However, to use these phenomenological models some assumptions have to be made that may influence the spiking profile of the modeled population of ANFs.

The peripheral auditory model presented in this work used a deterministic model of the ENI to obtain the most likely site of stimulation in every ANF. This assumption oversees the fact that ANFs do not initiate action potentials always in the same stimulation compartment, even when stimuli are identical. It is also shown that different stimulation sites are related to different spike latencies that affect the temporal fine structure of the spiking profile \citep{Javel2000}. Nevertheless, because the neural activity is effectively low-pass filtered with the temporal integration in the proposed model (see Eqs \ref{EqLt} and \ref{EqTIM}), it is expected that the categorical loudness at the final stage of the model is minimally affected by these changes in the fine structure.

Another limitation of the phenomenological model is that it counts with adaptation currents that are meant to emulate only the facilitation and refractoriness of the ANFs \citep{Joshi2017, Kipping2022}. It does not take into account the effect of adaptation and accommodation that influences the long-term spiking rate with electrical stimulation \citep{Zhang2007, Heffer2010}. This makes the phenomenological model ideal for short stimuli, but emulates poorly the spiking rate decrease in ANFs with longer stimulation periods. As shown in Figure 2 from \cite{Zhang2007}, adaptation appears within the first \(200~\text{ms}\) of stimulation and is more pronounced the higher the stimulation rate is. In the proposed computational model, the loudness index around MCL levels would be mainly driven by the spectral loudness summation with the exponential term in Eq \ref{EqIL}. Therefore, ignoring the adaptation and accommodation effects, mainly affected the performance of the model at low stimulation levels. This is a possible explanation for the discrepancies between simulations and data from \cite{Zhou2012} (see Figure \ref{figResultsE1}.A and Table \ref{tab_Exp1}), where the slope of normalized levels versus doubling the stimulation rate at THL levels decreased at stimulation rates higher than 1000 pps. \cite{Zhou2012} observed the opposite.

\subsection{Loudness through the auditory pathway}

The back-end of the proposed computational model presented four stages to go from peripheral neural activity to loudness perception, namely: 1) the neural excitation to loudness contribution, ii) the temporal integration, iii) the spatial integration, and iv) the loudness selection criterion (see Figure \ref{LoudModel}). Despite of no physiological claims done, it is clear that these stages are modeling the auditory pathway beyond the peripheral auditory system up to the cognition. The loudness selection criterion is a cognitive process and there is evidence of tonototy preserved up to the auditory cortex hinting that the spatial integration might be a cortical task as well \citep{Ress2013, Saenz2014}. The location of the temporal integration and the neural activity transformation into loudness contribution in the auditory pathway is a more uncertain topic.

The practical method and the DLM described by \cite{McKay2021} differ in wether the temporal integration should be implemented over the neural activity or the loudness contribution. Preliminary experiments with the proposed computational model showed that the loudness index increased in a disproportional way with higher stimulation rates if the temporal integration applied over the peripheral neural activity. This observations aligns better with locating the temporal integration in a more central location as in the practical method \citep{McKay2003}. Additionally, if the spectral loudness summation, which is modeled by the exponential factor in Eq \ref{EqExpfactor}, takes places in the cortex \citep{Rohl2011}, then the temporal integration also takes place in the cortex. However, the reality is that every stage in the auditory pathway consists of their own population of neurons, which at the same time could integrate incoming spike activity in many different ways. The temporal integration and the calculation of the loudness contribution in the proposed computational model could represent the cumulative effect of these physiological stages instead. 

In other implementations of the loudness models, two different temporal integration stages were combined \citep{Francart2014, Langner2020, McKay2021}. In this case, the temporal integration model from \cite{Oxenham1994} was applied after the loudness contribution calculation with a reduced equivalent rectangular duration (ERD) of 2~ms \citep{Francart2014}. In contrast, the implementation in our proposed model counts with an ERD of around 10~ms \citep{Oxenham2001}. The reduction was needed in order to incorporate the ``short-term loudness'' and ``long-term loudness'' calculations taken from the time-varying loudness (TVL) model presented by \cite{Glasberg2002}. However, the attack time of the automatic gain control circuit to obtain the short-term loudness has to be reduced to 1 ms as well. Both temporal models, the proposed by \cite{Oxenham1994} and by \cite{Glasberg2002}, were designed to account for the same psychoacoustic phenomena related to temporal resolution. For this reason, their parameters had to be fitted again, but also, it supports the idea of having multiple temporal models representing multiple stages in the auditory pathway. Nevertheless, there is still too much uncertainty about how perceptual features are transferred upwards in the auditory pathway to reach a definitive conclusion.

\subsection{Outlook}

The proposed loudness model offers a new perceptual feature based on the instantaneous specific loudness. This feature can be implemented in frameworks to simulate auditory experiments, i.e. \cite{Alvarez2023}, or algorithms to estimate the speech intelligibility index based on the peripheral neural activity, i.e. \cite{Alvarez2022}). It is also worth mentioning that the computational model could be improved to account for longer and time-variant stimuli by incorporating the TVL model as in \cite{Francart2014}.

Other possible future improvement is the extension of the phenomenological model of the ANF to account for the effects of accommodation and adaptation. This feature has been successfully included in recent phenomenological models \citep{Felsheim2024, deNobel2024}. Additionally, in order to improve the physiological aspects beyond the peripheral system, it might be beneficial to run studies based on objectives measures such as the auditory brainstem response (ABR), i.e. \cite{Stapells1997}, or the functional magnetic resonance imaging (fMRI), i.e. \cite{Behler2021}, and compare them to the predictions of the model. Those measures target the neural activity produced in upper stages of the auditory pathway.

As a final remark, since the loudness model for electric stimulation presented in this study is based on a temporal integration model originally fitted for acoustic stimuli, it could potentially be applied to acoustically stimulated peripheral models \citep{Bruce2018, Verhulst2018}. This might require a re-definition of the cochlear places to better represent the auditory filters that are, arguably, formed at the level of the basilar membrane \citep{Fletcher1933, Fletcher1940, Zwicker1957, moore1997}. 

\section{Conclusion}\label{sec13}

This work proposed a computational model to simulate loudness experiments in CI users. In the front-end, a deterministic 3D ENI was combined with a population of phenomenologically modeled ANFs. In the back-end, a loudness model was developed to return a categorical loudness level given the simulated peripheral neural activity. The model was capable to reproduce qualitatively the performance of real CI users in different loudness experiments and provides a new approach to calibrate peripheral auditory models in terms of stimulation levels. The proposed loudness model closes the gap between categorical loudness in real CI users and simulated neural activity. It also offers a new perceptual feature based on the peripheral neural activity.

\backmatter

\section*{Declarations}

\subsection*{Funding}

This work was supported in part by European Research Council (ERC) through European Union’s Horizon-ERC Program under Grant READIHEAR 101044753 – PI: WN, in part by the German Academic Exchange Service (Deutscher Akademischer Austauschdienst; DAAD) under Research Grants - Doctoral Programmes in Germany 2021/22, and in part by the DFG Cluster of Excellence under Grant EXC 2177/1 “Hearing4all”. (Corresponding author: Waldo Nogueira.)

\subsection*{Conflict of interest}

The authors declare that the research was conducted in the
absence of any commercial or ﬁnancial relationships that could be
construed as a potential conﬂict of interest.

\noindent


%%\bibliography{sn-bibliography}% common bib file

\end{document}